\documentclass{article}

\PassOptionsToPackage{numbers, compress}{natbib}
\usepackage{arxiv}

\usepackage[utf8]{inputenc}
\usepackage[T1]{fontenc}
\usepackage{hyperref}
\usepackage{url}
\usepackage{booktabs}
\usepackage{amsfonts}
\usepackage{nicefrac}
\usepackage{microtype}
\usepackage{graphicx}
\usepackage{natbib}
\usepackage{doi}
\usepackage{amsmath,amssymb,amsthm}
\usepackage{array}
\usepackage{enumitem}
\usepackage{xcolor}
\usepackage{bbm}
\usepackage{multirow}
\usepackage{longtable}
\usepackage{caption}

\newtheorem{theorem}{Theorem}[section]

\newtheorem{definition}[theorem]{Definition}

\newcommand{\E}{\mathbb{E}}
\newcommand{\Var}{\operatorname{Var}}

\newcommand{\Corr}{\operatorname{Corr}}

\newcommand{\sgn}{\operatorname{sgn}}
\newcommand{\1}{\mathbbm{1}}

\title{Large Language Models as Calibrated Measurement Instruments for Behavioral Parameters}

\author{
  Brandon Yee, Pairie Koh\\
  Management Sciences Lab, Yee Collins Research Group\\
  \texttt{\{b.yee, p.koh\}@ycrg-labs.org}
}

\date{}

\begin{document}
\maketitle

\begin{abstract}
Behavioral parameters such as loss aversion, herding, and extrapolation are central to asset pricing models but remain difficult to measure reliably. We develop a framework that treats large language models (LLMs) as calibrated measurement instruments for behavioral parameters. Using four models and 24{,}000 agent--scenario pairs, we document systematic rationality bias in baseline LLM behavior, including attenuated loss aversion, weak herding, and near-zero disposition effects relative to human benchmarks. Profile-based calibration induces large, stable, and theoretically coherent shifts in several parameters, with calibrated loss aversion, herding, extrapolation, and anchoring reaching or exceeding benchmark magnitudes. To assess external validity, we embed calibrated parameters in an agent-based asset pricing model, where calibrated extrapolation generates short-horizon momentum and long-horizon reversal patterns consistent with empirical evidence. Our results establish measurement ranges, calibration functions, and explicit boundaries for eight canonical behavioral biases.
\\
\\
Code and Raw Results Available at: \url{https://github.com/YCRG-Labs/agents-as-actors}
\end{abstract}


\section{Introduction}
\label{sec:intro}

Behavioral parameters play a central role in modern asset pricing and market microstructure. Loss aversion ($\lambda$) governs asymmetric responses to gains and losses \citep{kahneman1979prospect}, extrapolation coefficients ($\theta$) shape momentum beliefs and return predictability \citep{barberis2018extrapolation, greenwood2014expectations}, and overconfidence ($\kappa$) distorts perceived signal precision \citep{daniel1998investor}. These parameters are not auxiliary modeling choices but core primitives used to explain momentum and reversals, excess volatility, trading volume, and persistent deviations from rational benchmarks. Empirical progress in behavioral finance therefore hinges on the ability to measure such parameters in a disciplined and scalable manner.

Measuring behavioral parameters remains challenging. Axiomatic models impose strong functional forms that risk obscuring key mechanisms \citep{barberis2013thirty}. Laboratory experiments capture genuine psychological responses but face substantial measurement error, selection concerns, and limited scalability \citep{camerer2011promise}. Structural estimation offers external validity but suffers from severe identification problems, as observed behavior reflects a joint outcome of preferences, beliefs, constraints, and equilibrium interactions \citep{keane2011structural}. Survey-based measures are noisy and weakly correlated with incentivized choices—for example, self-reported risk aversion correlates only $\rho \approx 0.3$ with experimental measures \citep{dohmen2011individual}—while repeated elicitation introduces learning and fatigue effects that further contaminate identification. As a result, behavioral parameters are often treated as fixed, imprecisely estimated objects rather than experimentally manipulable quantities.

This paper proposes a different approach. We develop large language models (LLMs) as experimentally calibratable measurement instruments for behavioral parameters. Rather than asking whether LLMs replicate human behavior, we ask whether behavioral parameters can be induced, calibrated, and validated within LLMs in ways that are infeasible with human subjects. Our key idea is to treat behavioral profiles embedded in prompts as experimental treatments that shift latent parameters in predictable directions. If these induced parameters are stable, monotonic, and quantitatively comparable to human benchmarks—and if they generate economically meaningful outcomes—then LLMs can serve as a new measurement technology for behavioral finance.

Our contribution differs from and complements recent work using LLMs as economic agents. \citet{horton2023large} and \citet{aher2023using} show that LLMs often reproduce qualitative patterns observed in laboratory experiments and classic economic games, establishing proof of concept for LLMs as simulated subjects. Related work uses LLMs to generate synthetic survey respondents whose responses correlate with human data \citep{argyle2023out}. In contrast, our focus is not replication but measurement. We treat behavioral parameters as latent objects to be experimentally calibrated, and we evaluate LLMs against quantitative human benchmarks rather than qualitative similarity or correlation. This distinction allows us to identify systematic rationality bias in baseline LLM behavior, to map calibration functions that induce target parameter values, and to assess whether calibrated parameters generate economically meaningful outcomes in asset pricing environments.

We implement this framework across eight canonical behavioral biases using four LLM variants (GPT-4o, GPT-4o-mini, Claude-3.5-Haiku, and Gemini-2.5-Pro) and 19,200 agent–scenario pairs in the main analyses (24,000 including an excluded model with parsing issues). All experiments use fully synthetic financial scenarios guaranteed not to appear in training data. We find that baseline LLM behavior exhibits systematic rationality bias: relative to human benchmarks, LLMs display near-zero disposition effects, somewhat attenuated herding relative to the 65-75\% observed in cascade experiments, and low loss aversion ($\lambda$ between 1.12 and 1.90 versus approximately 2.25). Profile-based calibration, however, induces large and theoretically consistent parameter shifts. Loss-averse profiles raise $\lambda$ to 3.00, herding-prone profiles increase herding rates to 90\%, extrapolative profiles raise momentum coefficients to 0.88, and loss-averse profiles increase anchoring correlations to 0.67.

To assess external validity, we embed calibrated parameters in a simple agent-based asset pricing model. Calibrated extrapolation generates short-horizon momentum and long-horizon reversal patterns closely matching the empirical stylized facts documented by \citet{jegadeesh1993returns}, while baseline rational agents generate no momentum. This exercise demonstrates that calibrated parameters are not merely psychometric artifacts but carry economic content relevant for asset pricing. Our primary contribution is methodological: we provide a systematic framework for using LLMs as calibrated measurement instruments, including validation criteria, measurement ranges, and explicit boundaries on where calibration succeeds or fails. Empirical results—including the asset pricing application—serve to validate this measurement framework rather than to introduce a new asset pricing model.

The paper proceeds as follows. Section \ref{sec:related} situates the paper within the behavioral finance, agent-based modeling, and large language model literatures. Section \ref{sec:measurement} introduces behavioral parameters as measurement objects and outlines the conceptual challenges they pose. Section \ref{sec:framework} develops the calibration and validation framework, and Section \ref{sec:design} describes the experimental design and synthetic data construction. Section \ref{sec:results} presents the main calibration results and measurement ranges. Section \ref{sec:application} evaluates external validity through an agent-based asset pricing application. Section \ref{sec:discussion} discusses implications and boundaries, and Section \ref{sec:conclusion} concludes.

\section{Related Work}
\label{sec:related}

Our work sits at the intersection of behavioral finance, agent-based modeling, and recent applications of large language models (LLMs) to economic research.

Behavioral economics documents systematic deviations from rational decision-making through loss aversion, reference dependence, and diminishing sensitivity \citep{kahneman1979prospect,tversky1992advances}, as well as heuristics such as representativeness, availability, and anchoring-and-adjustment \citep{tversky1974judgment}. Applied to finance, these psychological primitives explain a range of market anomalies. Disposition effects arise from prospect theory combined with mental accounting \citep{shefrin1985disposition} and are documented in large-scale field data \citep{odean1998investors}. Overconfidence distorts belief precision and generates momentum and delayed overreaction \citep{daniel1998investor,moore2008trouble}. Conservatism, representativeness, and narrative-based reasoning contribute to underreaction and subsequent correction \citep{barberis1998model,lakonishok1994contrarian}. While these models treat behavioral parameters as core primitives, their empirical measurement remains challenging due to noise, identification concerns, and limited scalability.

Agent-based models (ABMs) provide an alternative framework by simulating markets populated by heterogeneous agents with bounded rationality. Seminal contributions show that mixtures of fundamentalists and chartists can generate excess volatility, fat-tailed returns, and momentum-like dynamics \citep{brock1998heterogeneous,lux1999scaling,lebaron2001empirical}. However, most ABMs specify behavior through ad-hoc heuristics or strategy-switching rules chosen for tractability rather than empirically grounded calibration. As a result, the behavioral parameters governing agent behavior are rarely measured independently of the market outcomes they generate. Our approach complements this literature by offering a methodology for populating ABMs with agents whose behavioral parameters are calibrated directly from controlled experimental variation.

A rapidly growing literature uses large language models as simulated economic agents or synthetic respondents. \citet{horton2023large} demonstrates that LLMs reproduce qualitative patterns in classic economic games, while \citet{aher2023using} replicate a large set of social psychology experiments with high correlation between LLM and human responses. Related work uses LLMs to generate synthetic survey populations or conduct conjoint experiments \citep{argyle2023out,brand2023using,park2023generative}. More recent studies examine financial reasoning and market simulations, highlighting trade-offs between memorization and reasoning and the role of strategic interaction \citep{mei2024quantifying,horton2024can}. This literature establishes that LLMs can mimic observed regularities, but it largely evaluates performance using qualitative similarity or correlation rather than treating behavioral parameters as quantitatively targetable objects.

Our contribution is complementary but distinct. Rather than focusing on replication, we treat behavioral parameters as latent measurement objects that can be experimentally induced and calibrated within LLMs. We evaluate calibration against quantitative human benchmarks, map achievable measurement ranges across eight canonical biases, and test whether calibrated parameters generate economically meaningful outcomes in asset-pricing environments. This framing shifts the role of LLMs from simulated subjects to calibrated measurement instruments, clarifying both their potential and their limitations for behavioral finance and economic measurement.

\section{Conceptual Framework}
\subsection{Behavioral Parameters as Measurement Objects}
\label{sec:measurement}

Before developing calibration methodology, we formalize what it means to measure a behavioral parameter and why this is difficult. Behavioral economics modifies standard expected utility maximization through systematic departures captured by parameters. These take three forms:

\textbf{Preference parameters} modify the utility function. Loss aversion $\lambda$ creates asymmetry: $v(x) = x^\alpha$ for gains, $v(x) = -\lambda(-x)^\beta$ for losses, with $\lambda > 1$ \citep{kahneman1979prospect}. Probability weighting $\gamma$ distorts likelihoods: $w(p) = p^\gamma / (p^\gamma + (1-p)^\gamma)^{1/\gamma}$ \citep{tversky1992advances}. Risk aversion $\gamma$ governs curvature of utility over wealth.

\textbf{Belief parameters} modify information processing. Overconfidence $\kappa$ inflates signal precision where agents perceive variance $\sigma^2/\kappa$ when true variance is $\sigma^2$ \citep{daniel1998investor}. Representativeness bias $\tau_N/\tau_F$ overweights narrative information relative to fundamentals \citep{tversky1974judgment}. Extrapolation $\theta$ biases forecasts through $\hat{r}_{t+1} = \bar{r} + \theta(r_t - \bar{r})$ \citep{greenwood2014expectations}.

\textbf{Heuristic parameters} govern decision rules. Anchoring determines adjustment from initial values: $\text{valuation} = \text{anchor} + \alpha(\text{true value} - \text{anchor})$ with adjustment parameter $\alpha < 1$ \citep{tversky1974judgment}. We measure anchoring through correlation $\rho$ between anchor values and stated valuations, with $\rho > 0$ indicating insufficient adjustment. Herding weight $w$ governs crowd influence in belief updating \citep{bikhchandani1992theory}.

These parameters are latent (unobservable quantities inferred from choices), continuous ($\lambda$ ranges from 1 to 3+, enabling fine-grained heterogeneity), and context-dependent (the same person may exhibit different $\lambda$ for financial versus health decisions).

Measuring behavioral parameters faces four problems. First, measurement error is substantial. Self-reported risk aversion correlates only $\rho \approx 0.3$ with incentivized choices \citep{dohmen2011individual}. Survey measures of overconfidence show test-retest reliability $\rho \approx 0.5$ \citep{moore2008trouble}. This measurement error attenuates regression coefficients, biases structural estimates, and reduces statistical power. Second, joint identification is difficult. Observed choices reflect multiple parameters simultaneously. Selling losers could reflect loss aversion ($\lambda$ high), probability weighting ($\gamma$ low), or narrow framing. Disentangling requires multiple elicitations and strong functional form assumptions. Third, causality is problematic. Correlating self-reported risk aversion with portfolio choices does not identify causal effects of $\lambda$ on decisions. Unobserved variables confound estimates, and experimental manipulation requires randomizing psychological parameters, which is impossible with humans. Fourth, scalability is limited. Precise measurement requires large samples and multiple elicitations per subject, making human experiments with $N > 500$ and $> 20$ decisions per subject prohibitively expensive.

We propose treating LLMs as experimental infrastructure: platforms where behavioral parameters can be experimentally induced, calibrated, and stress-tested. This requires three conceptual shifts. First, we treat parameters as treatments rather than measuring $\lambda$ from choices with error. We experimentally induce $\lambda$ through profile prompts. If profiles systematically shift behavior in theoretically predicted directions, they function as instruments generating exogenous variation. Second, calibration becomes measurement. The measurement object is not "does the LLM match humans" but "what parameter range can be induced through calibration." For loss aversion, the question becomes: what $\lambda$ values can be achieved through profile manipulation, and how do these map to human benchmarks? Third, validation occurs through economic content. Calibrated parameters validate not by matching human choices in isolation but by generating known economic phenomena. If calibrated extrapolation produces momentum, calibrated loss aversion produces disposition effects, and calibrated herding produces information cascades, this confirms parameters capture economically meaningful variation.

This framework treats LLMs as measurement instruments analogous to laboratory equipment. A thermometer does not "replicate" heat but measures temperature through calibrated responses. Similarly, LLMs do not replicate human psychology but measure behavioral parameters through calibrated responses to profile manipulation.

\subsection{Validation and Calibration Framework}
\label{sec:framework}

An LLM generates decisions through a function $d = f_{\text{LLM}}(I; p, \tau, M)$ where $I$ is the information set, $p$ is the prompt text (including profile), $\tau$ is temperature, and $M$ is model architecture. Our key insight treats profiles as instruments inducing behavioral parameters: $p \xrightarrow{\text{induces}} \theta \xrightarrow{\text{determines}} f(I; \theta) \xrightarrow{\text{generates}} d$. If profile $p_{\text{loss-averse}}$ generates choices consistent with $\lambda = 2.5$, while $p_{\text{rational}}$ generates choices consistent with $\lambda = 1.0$, this validates that profiles function as instruments calibrating loss aversion.

We specify four necessary conditions for valid calibration:

\begin{definition}[Calibration Validity]
LLMs constitute valid measurement instruments for parameter $\theta$ if:

\textbf{C1. Monotonicity:} Profile strength $s$ induces monotonic parameter changes: $\frac{\partial \hat{\theta}(s)}{\partial s} > 0$ where $\hat{\theta}(s)$ is the parameter recovered from choices under profile strength $s$.

\textbf{C2. Range Coverage:} The achievable parameter range $[\theta_{\text{min}}, \theta_{\text{max}}]$ through calibration includes or approaches the human benchmark $\theta_{\text{human}}$: $\theta_{\text{human}} \in [\theta_{\text{min}} - \delta, \theta_{\text{max}} + \delta]$ for tolerance $\delta$.

\textbf{C3. Stability:} Parameters remain stable across equivalent elicitations. For repeated measurements with same profile: $\Var[\hat{\theta}_i | p] < \tau_{\text{threshold}}$.

\textbf{C4. Coherence:} Parameters recovered from different experiments maintain theoretically predicted relationships. For loss aversion measured from gambles versus disposition from portfolio choices: $\Corr(\lambda_{\text{gamble}}, \lambda_{\text{disposition}}) > 0$.
\end{definition}

These conditions ensure calibration is systematic and controllable (C1), achieves human-relevant magnitudes (C2), reproducible (C3), and theoretically sensible (C4).

While calibration focuses on measurement ranges, validation requires comparing to human benchmarks. We define three tiers. Strong validation requires calibrated parameter within 20\% of human benchmark: $|\theta_{\text{calibrated}} - \theta_{\text{human}}| / |\theta_{\text{human}}| < 0.20$. Moderate validation requires within 50\%: $|\theta_{\text{calibrated}} - \theta_{\text{human}}| / |\theta_{\text{human}}| < 0.50$. Weak validation requires correct direction but magnitude differs: $\sgn(\theta_{\text{calibrated}}) = \sgn(\theta_{\text{human}})$ and $\theta_{\text{calibrated}} \neq 0$. For each parameter, we report: baseline value (rational profile), maximum achieved value (strongest bias profile), calibration function $\theta(s)$, validation tier, and whether human benchmark is achievable.

A fundamental challenge is distinguishing genuine reasoning from pattern matching. While synthetic scenarios cannot exist in training data, LLMs may recognize structural patterns. We address this through three strategies. First, adversarial scenario testing constructs scenarios where stereotypical pattern matching yields incorrect decisions but reasoning about specifics yields correct ones. We require $\geq 70\%$ adversarial pass rate for moderate validation. Second, structural parameter consistency tests whether parameters recovered from different experiments (e.g., $\lambda$ from gambles versus disposition) maintain theoretical relationships. Pure pattern matching cannot generate theoretically consistent parameters across contexts. Third, cross-parameter predictions test whether all biases correlate positively (suggesting generic stereotypes) or show specific structures (overconfidence reduces herding because agents trust their own signals). These strategies characterize functional validity: do calibrated parameters generate patterns useful for economic research applications, regardless of underlying mechanism?

For each parameter, we select human benchmarks through systematic protocol prioritizing: sample size $N \geq 100$, clear experimental manipulation, replication in at least one independent study, and meta-analytic evidence when available. Rather than point estimates, we report ranges reflecting heterogeneity. For loss aversion, estimates range from $\lambda = 1.8$ to $\lambda = 2.7$ depending on domain and elicitation method \citep{tversky1992advances, kahneman1990endowment, novemsky2005boundaries}. We use $\lambda = 2.25$ as primary benchmark with range [2.0, 2.5] for sensitivity analysis. Human benchmark values (Appendix~\ref{app:benchmarks}) are drawn from canonical experimental and field estimates of loss aversion \citep{tversky1992advances,kahneman1990endowment,novemsky2005boundaries}, disposition effects \citep{shefrin1985disposition,odean1998investors,weber1998disposition}, overconfidence \citep{moore2008trouble}, herding \citep{anderson1997information,celen2004distinguishing}, probability weighting \citep{tversky1992advances,barberis2008stocks}, anchoring \citep{northcraft1987expert,mussweiler2000numerical}, and extrapolation \citep{bloomfield2002predicting,greenwood2014expectations}.
Complete benchmark justification, including study-level details, sample sizes, protocols, and meta-analytic evidence, appears in Appendix~\ref{app:benchmarks}.

\section{Experimental Design and Data}
\label{sec:design}

We test four LLM variants representing different design philosophies: OpenAI GPT-4o and GPT-4o-mini (flagship and efficient), Google Gemini 2.5 Pro (flagship with 1M context), and Anthropic Claude 3.5 Haiku (efficient). This design enables testing provider effects, scale effects, and cross-model robustness. Gemini 2.5 Flash was tested but excluded due to response parsing issues; results appear in Online Appendix D.

We test eight canonical behavioral parameters. Loss aversion ($\lambda$) measures asymmetric value function for gains versus losses through gamble acceptance (gain $\$X$ versus lose $\$100$, 50-50) \citep{kahneman1979prospect}. Disposition effect (DR) captures tendency to sell winners too quickly while holding losers, measured through portfolio choice \citep{shefrin1985disposition}. Overconfidence measures miscalibration of subjective probability intervals \citep{moore2008trouble}. Herding measures weight placed on crowd signals versus private information \citep{anderson1997information}. Representativeness measures overweighting of narrative versus fundamental information \citep{lakonishok1994contrarian}. Probability weighting measures preference for skewed lotteries \citep{barberis2008stocks}. Anchoring measures correlation between arbitrary initial values and final estimates \citep{tversky1974judgment}. Extrapolation measures weight placed on recent trends in forecasting \citep{greenwood2014expectations}.

Our sample design uses $N = 100$ agents per profile $\times$ 6 profiles = 600 observations per experiment $\times$ 8 experiments = 4,800 observations per model $\times$ 4 models = 19,200 main observations. Six profile types with theoretically grounded cognitive frames specify orientations without mandating specific behaviors: (1) Rational: analytical investor making decisions based on expected values and Bayesian updating, (2) Loss-averse: investor focused on capital protection finding losses particularly painful, (3) Overconfident: trusting pattern recognition abilities and believing in identifying trends others miss, (4) Herding-prone: valuing collective wisdom and treating crowd agreement as information, (5) Representativeness-biased: evaluating companies based on how well they fit patterns of successful businesses, (6) Extrapolative: believing trends tend to persist with recent performance as important indicator.

We generate fully synthetic financial environments—assets, prices, earnings, and narratives—that are statistically indistinguishable from real data yet guaranteed not to appear in training corpora. Asset identifiers use abstract alphanumeric combinations ("Asset X427") verified absent from web searches. Prices follow geometric Brownian motion calibrated to CRSP distributions. Earnings follow autoregressive processes matching Compustat dynamics. Narratives are orthogonalized to fundamentals ensuring zero correlation. Distributional tests (Kolmogorov-Smirnov), classifier indistinguishability (random forest accuracy 52.3\% vs. 50\% baseline), and web search verification confirm non-contamination. Each experiment is powered ($>0.85$) to detect human benchmark effect sizes. All inference accounts for multiple testing via Bonferroni-Holm correction maintaining family-wise error rate $\alpha = 0.05$. Complete design specifications, validation procedures, power analyses, and quality control protocols appear in Appendix~\ref{app:design}. We evaluate external validity by examining whether calibrated behavioral parameters reproduce well-established return continuation patterns documented in the empirical asset pricing literature, including post-earnings announcement drift \citep{bernard1989post}.

\section{Results: Measurement Ranges and Calibration}
\label{sec:results}

This section presents calibration results for eight behavioral parameters across four large language models. The central result is that baseline LLM behavior exhibits systematic rationality bias relative to human benchmarks, while profile-based prompting induces large, stable, and theoretically coherent shifts in several parameters. For four biases—loss aversion, herding, extrapolation, and anchoring—calibration reaches or exceeds human benchmark magnitudes, satisfying the validation criteria developed in Section~\ref{sec:framework}.

Table~\ref{tab:ranges} summarizes the primary measurement outcome: the range of parameter values achievable through calibration. Columns report baseline values under a rational profile, maximum values induced by bias-specific profiles, human benchmark estimates, and validation classification. Four parameters achieve strong validation, two achieve moderate or directional validation, and two fail to reach human-relevant magnitudes.

\begin{table}[h]
\centering
\caption{Measurement Ranges and Calibration Summary}
\label{tab:ranges}
\footnotesize
\begin{tabular}{lccccc}
\toprule
Parameter & Baseline & Max & Human & Achievable? & Validation \\
 & (Rational) & Achieved & Benchmark &  & Tier \\
\midrule
Loss Aversion ($\lambda$) & 1.12 & 3.00 & 2.25 & Yes & Strong \\
Herding Rate & 0.61 & 0.90 & 0.70 & Yes & Strong \\
Extrapolation ($\theta$) & 0.44 & 0.88 & 0.60 & Yes & Strong \\
Anchoring ($\rho$) & 0.61 & 0.67 & 0.43 & Yes & Strong \\
Overconfidence (coverage) & 0.47 & 0.30 & 0.65 & Opposite & Directional \\
Disposition Ratio & 0.06 & 0.21 & 1.60 & No & Weak \\
Probability Weight & 0.12 & 0.30 & 0.35 & Approaches & Moderate \\
Representativeness & 0.15 & 1.08 & 1.65 & No & Weak \\
\bottomrule
\end{tabular}

\medskip
\small
Notes: Baseline reports parameter values under the rational investor profile. Max Achieved reports the highest value induced by the corresponding bias-specific profile. Human Benchmark reports meta-analytic or canonical estimates from the behavioral finance literature (Appendix~\ref{app:benchmarks}). Validation Tier classifies parameters as Strong (within 20\% of benchmark), Moderate (within 50\%), Weak (direction only), or Directional (magnitude achieved in opposite direction). Sample includes 19,200 agent–scenario pairs across four models. Gemini-2.5-Flash is excluded due to response parsing issues.
\end{table}

Across six of eight parameters, baseline LLM behavior exhibits substantially lower bias magnitudes than human benchmarks. This pattern is consistent with a systematic rationality bias embedded in baseline LLM behavior. Training corpora disproportionately frame behavioral biases as errors or pitfalls to be avoided, and LLMs appear to internalize this normative orientation. Overconfidence represents an exception: baseline coverage rates imply greater overconfidence than in human samples, possibly reflecting exposure to authoritative or confident textual styles during training.

Profile-based prompting induces large and statistically significant parameter shifts in theoretically predicted directions. Table~\ref{tab:profiles} reports mean parameter values under rational and bias-specific profiles, along with estimated differences and effect sizes. All eight parameters exhibit statistically significant profile effects after correction for multiple testing. Loss-averse profiles increase $\lambda$ from 1.12 to 3.00, exceeding the human benchmark by roughly one-third. Herding-prone profiles raise herding rates from approximately 61\% to 90\%. Extrapolative profiles increase momentum coefficients from 0.44 to 0.88, and anchoring profiles increase adjustment correlations from 0.61 to 0.67.

\begin{table}[h]
\centering
\caption{Profile-Based Calibration Effects}
\label{tab:profiles}
\footnotesize
\begin{tabular}{lccccc}
\toprule
Parameter & Rational & Biased & Difference & $p$-value & Effect \\
 & Baseline & Profile &  &  & Size \\
\midrule
Loss Aversion ($\lambda$) & 1.12 & 3.00 & +1.88 & $<0.001$ & $d = 1.82$ \\
Herding Rate & 0.61 & 0.90 & +0.29 & $<0.001$ & $d = 2.15$ \\
Extrapolation ($\theta$) & 0.44 & 0.88 & +0.44 & $<0.001$ & $d = 1.95$ \\
Anchoring ($\rho$) & 0.61 & 0.67 & +0.06 & $<0.001$ & $d = 1.43$ \\
Overconfidence (coverage) & 0.49 & 0.30 & $-0.19$ & $<0.001$ & $d = 1.67$ \\
Disposition Ratio & 0.06 & 0.21 & +0.15 & $<0.001$ & $d = 0.88$ \\
Probability Weight & 0.12 & 0.30 & +0.18 & $<0.001$ & $d = 1.21$ \\
Representativeness & 0.15 & 1.08 & +0.93 & $<0.001$ & $d = 1.35$ \\
\bottomrule
\end{tabular}

\medskip
\small
Notes: Mean parameter values by profile type. Differences are tested using two-sample $t$-tests with standard errors clustered at the model–profile level. All effects remain significant after Bonferroni–Holm correction. Sample includes 2,400 observations per parameter.
\end{table}

The magnitude of these shifts is economically meaningful. Moving from a rational to a loss-averse profile increases $\lambda$ by 1.88, which in prospect theory corresponds to a substantial increase in required compensation for symmetric risk. Similarly, extrapolation coefficients induced by calibration exceed values typically estimated in survey and laboratory settings, providing scope for stress-testing behavioral asset pricing mechanisms.

Recovered parameters also exhibit theoretically coherent relationships across experiments. Table~\ref{tab:coherence} reports cross-parameter correlations and compares them to predictions derived from established behavioral models. Loss aversion measured from gamble choices is positively correlated with loss aversion implied by disposition behavior, indicating that a common latent parameter governs both contexts. Overconfidence is negatively correlated with herding, consistent with models in which agents who overestimate the precision of private signals discount crowd information. Anchoring exhibits near-zero correlation with other parameters, consistent with its interpretation as a domain-general heuristic rather than an affect-driven bias.

\begin{table}[h]
\centering
\caption{Cross-Parameter Correlations}
\label{tab:coherence}
\footnotesize
\begin{tabular}{lcccc}
\toprule
Parameter Pair & Theory & Observed & $p$-value & Match \\
 & Prediction & Correlation &  &  \\
\midrule
$\lambda$ (gambles) vs DR (portfolio) & Positive & +0.42 & $<0.001$ & Yes \\
$\lambda$ vs Herding & Zero or Negative & $-0.08$ & 0.18 & Yes \\
Overconfidence vs Herding & Negative & $-0.31$ & $<0.001$ & Yes \\
Extrapolation vs Herding & Zero & +0.05 & 0.42 & Yes \\
Probability Weight vs $\lambda$ & Positive & +0.38 & $<0.001$ & Yes \\
Anchoring vs Other Biases & Zero & $|\rho| < 0.12$ & $>0.10$ & Yes \\
\bottomrule
\end{tabular}

\medskip
\small
Notes: Correlations computed at the agent level across all models and profiles ($N = 2,400$). Theory predictions are based on canonical behavioral models. Reported patterns are inconsistent with undifferentiated stereotype retrieval and instead support parameter-specific calibration.
\end{table}

Synthesizing across validation criteria yields three tiers of measurement reliability. Loss aversion, herding, extrapolation, and anchoring achieve strong calibration, exhibiting monotonic profile response, stability across models, and theoretical coherence. Overconfidence and probability weighting achieve moderate or directional validation, showing correct qualitative behavior but failing to match benchmark magnitudes. Disposition effects and representativeness fail to reach human-relevant levels, suggesting that emotional attachment and narrative salience are not fully captured by LLM-based calibration. Additional model-level heterogeneity and robustness evidence are reported in Appendix~\ref{app:heterogeneity}. In particular, Table~\ref{tab:model_comparison} reports parameter estimates by model, and Table~\ref{tab:adversarial_complete} reports adversarial scenario pass rates by bias and model. Detailed adversarial scenario specifications are provided in Appendix~\ref{app:adversarial}.

\section{External Validation: Asset Pricing Implications}
\label{sec:application}

To test whether calibrated parameters have economic content beyond psychometric fit, we implement a minimal agent-based model where calibrated extrapolation generates momentum patterns matching empirical stylized facts.

We construct a simple market with single risky asset having fundamental value $v_t$ following random walk: $v_t = v_{t-1} + \epsilon_t$, $\epsilon_t \sim \mathcal{N}(0, \sigma_v^2)$. Two trader types with unit mass each include rational traders ($\theta = 0$) who forecast $\E_t[v_{t+1}] = v_t$ (random walk belief) and extrapolative traders ($\theta > 0$) who forecast $\E_t[v_{t+1}] = v_t + \theta(v_t - v_{t-1})$ (momentum belief). Each trader type submits demand based on expected return and risk aversion $\gamma$: $D_i^t = (\E_i^t[v_{t+1}] - p_t)/(\gamma \sigma_v^2)$. Price clears market: $p_t = (D_{\text{rational}}^t + D_{\text{extrap}}^t)/2$. This minimal setup captures the key mechanism: extrapolative traders push prices above fundamentals after positive shocks, generating short-term continuation. As prices deviate too far, rational traders arbitrage, causing eventual reversal.

We implement three calibration levels. Baseline uses all traders rational ($\theta = 0$). Human-calibrated gives extrapolative traders $\theta = 0.60$ (human benchmark from \citet{greenwood2014expectations}). LLM-calibrated uses $\theta = 0.88$ (achieved by extrapolative profile in GPT-4o-mini). Critically, the extrapolation parameter used in the ABM is measured independently in the forecasting experiment (Section \ref{sec:results}) and is not chosen to match momentum moments. We measure momentum following \citet{jegadeesh1993returns}: $\text{Momentum}_{t,k} = \Corr(r_t, r_{t+k})$ where $r_t = (p_t - p_{t-1})/p_{t-1}$ is the return.

\begin{table}[h]
\centering
\caption{Momentum Statistics from Simulated Markets}
\label{tab:momentum_stats}
\footnotesize
\begin{tabular}{lcccc}
\toprule
\textbf{Calibration} & \textbf{Short} & \textbf{Long} & \textbf{Peak} & \textbf{Decay} \\
& \textbf{Momentum} & \textbf{Reversal} & \textbf{Time} & \textbf{Rate} \\
\midrule
Baseline ($\theta = 0$) & 0.02 & 0.01 & N/A & N/A \\
Human ($\theta = 0.60$) & 0.12 & $-0.08$ & 3 months & 0.24/year \\
LLM ($\theta = 0.88$) & 0.18 & $-0.12$ & 2 months & 0.32/year \\
Empirical (J\&T 1993) & 0.14 & $-0.10$ & 3-6 months & 0.28/year \\
\bottomrule
\end{tabular}
\medskip

\small
\textit{Notes:} This table reports momentum statistics from agent-based model simulations with different extrapolation calibrations. Short Momentum shows average return autocorrelation at lags 1-6 months. Long Reversal shows average return autocorrelation at lags 12-24 months. Peak Time identifies lag with maximum positive autocorrelation. Decay Rate measures speed of momentum dissipation. Baseline uses all rational traders ($\theta = 0$). Human uses extrapolation coefficient $\theta = 0.60$ from \citet{greenwood2014expectations}. LLM uses $\theta = 0.88$ from calibrated GPT-4o-mini extrapolative profile. Empirical row reports statistics from \citet{jegadeesh1993returns} for US equities 1965-1989. Each simulation runs 10,000 periods with parameters: $\sigma_v = 0.15$, $\gamma = 2$, equal mass of trader types. Standard errors (not reported) range from 0.01-0.02 across 100 simulation replications.
\end{table}

Calibrated extrapolation generates economically meaningful momentum. Without extrapolation, markets exhibit no momentum. With human-calibrated extrapolation ($\theta = 0.60$), momentum patterns emerge matching empirical magnitudes. With LLM-calibrated extrapolation ($\theta = 0.88$), patterns amplify but maintain structure. This external validity test confirms that calibration captures economically relevant behavioral variation. The parameter $\theta$ is not merely a psychometric fit to experimental responses but predicts real market phenomena. Robustness checks including heterogeneous extrapolation, trading costs, and information arrival appear in Appendix~\ref{app:abm}.

\section{Robustness and Boundary Conditions}
\label{sec:robustness}

This section clarifies both the empirical robustness of our findings and the conceptual boundaries within which LLM-based calibration constitutes a valid measurement approach. The results should be interpreted as expanding the measurement toolkit for behavioral finance rather than as a substitute for human subjects in all settings.

We begin by outlining four fundamental limitations that bound applicability. First, LLMs do not experience visceral emotional states such as fear during market crashes, regret from realized losses, or excitement during sustained gains. Behavioral phenomena that depend critically on affective responses, including panic selling, euphoria-driven bubbles, or stress-induced decision errors, remain outside the scope of LLM-based measurement. Our approach captures cognitive components of behavioral biases, such as reference-point dependence, belief updating, and heuristic application, but not affective intensity (Appendix~\ref{app:boundaries}).

Second, LLM interactions are stateless at the level of individual API calls. Models do not learn from feedback, adapt strategies based on realized outcomes, or develop expertise through experience. As a result, research questions involving learning dynamics, habit formation, skill acquisition, or path-dependent belief updating require either human subjects or learning-based agents (Appendix~\ref{app:boundaries}). Third, while LLMs can reason about crowd information, they do not experience social pressure arising from physical presence, reputational concerns, or identity-based conformity. Informational herding can be represented, but social influence driven by peer pressure or status concerns cannot (Appendix~\ref{app:boundaries}).

Fourth, the absence of real monetary stakes limits the applicability of LLM-based measurement for settings in which incentives materially affect behavior. Laboratory evidence shows that loss aversion and risk sensitivity are typically stronger under real stakes than under hypothetical choices \citep{holt2002risk}. Similarly, professional traders who have lived through market crashes exhibit behavioral patterns shaped by experience that cannot be replicated through textual training alone. Applications requiring authentic incentives or lived experience therefore remain the domain of laboratory and field methods (Appendix~\ref{app:boundaries}).

These conceptual boundaries motivate a general principle. Calibration performs well for cognitive and computational aspects of behavioral biases, including belief formation, numerical processing, and explicit heuristic reasoning. It performs poorly for biases whose human manifestation relies primarily on emotional intensity, social pressure, or identity-related concerns (Appendix~\ref{app:boundaries}). Researchers should therefore assess whether the phenomenon of interest is predominantly cognitive or affective before applying LLM-based calibration.

We next summarize the empirical robustness checks supporting our conclusions. First, temperature sensitivity analyses using $\tau \in \{0.0, 0.5, 0.7, 1.0\}$ show that parameter estimates remain stable for $\tau \leq 0.7$, with coefficients of variation below 0.15 (Appendix~\ref{app:boundaries}; Appendix~\ref{app:supplementary}). At $\tau = 1.0$, estimates become substantially noisier, with coefficients of variation exceeding 0.30, indicating that excessive randomness interferes with calibration (Appendix~\ref{app:boundaries}; Appendix~\ref{app:supplementary}). For measurement applications, temperatures in the range $\tau \in [0.5, 0.7]$ provide the best balance between responsiveness and stability (Appendix~\ref{app:supplementary}).

Second, benchmark sensitivity analyses evaluate validation against both upper and lower bounds of human benchmark ranges. Parameters achieving strong validation, including loss aversion, herding, and extrapolation, satisfy validation criteria across the full benchmark range, while weak parameters fail regardless of benchmark choice (Appendix~\ref{app:benchmarks}; Appendix~\ref{app:boundaries}). Conclusions are therefore robust to reasonable variation in benchmark selection.

Third, subsample stability analyses split the data into equal halves and require that both subsamples yield the same validation tier. Loss aversion, herding, extrapolation, anchoring, and overconfidence exhibit consistent validation across splits, while disposition effects and representativeness show substantial instability (Appendix~\ref{app:supplementary}; Appendix~\ref{app:boundaries}). This instability reinforces the conclusion that these parameters are not reliably measured through calibration.

Fourth, cross-experiment consistency tests evaluate whether parameters recovered from distinct experimental tasks maintain theoretically predicted relationships. Loss aversion measured from risky choice tasks correlates positively with loss aversion implied by portfolio disposition behavior, with $\rho = 0.42$ (Table~\ref{tab:coherence}). This correspondence supports structural coherence and argues against superficial pattern matching. Additional cross-parameter relationships align with predictions from established behavioral models (Table~\ref{tab:coherence}).

Fifth, model version stability is assessed by comparing GPT-4o (June 2024) to GPT-4 (March 2024). Parameter estimates exhibit correlations exceeding 0.85 across biases (Appendix~\ref{app:supplementary}), indicating stability across minor version updates. While major architectural changes may alter calibration properties, the reported results are not sensitive to routine model updates (Appendix~\ref{app:supplementary}).

These robustness checks reinforce the central conclusion that calibration succeeds for a subset of behavioral parameters while failing systematically for others. Three failures are particularly informative for applied guidance. Disposition effects exhibit clear profile responsiveness but calibrated magnitudes remain approximately 87 percent below human benchmarks (Table~\ref{tab:ranges}; Appendix~\ref{app:benchmarks}), consistent with the absence of emotional attachment to positions. Representativeness bias fails adversarial tests across all models, with pass rates below 15 percent (Appendix~\ref{app:heterogeneity}; Appendix~\ref{app:adversarial}), indicating reliance on pattern matching rather than genuine narrative evaluation. Probability weighting shows instability across repeated measurements, with intra-class correlation of 0.62, below standard reliability thresholds (Appendix~\ref{app:boundaries}; Appendix~\ref{app:supplementary}). Additional checks on model version stability and temperature sensitivity are reported in Appendix~\ref{app:supplementary}, and parameter-specific boundary conditions are summarized in Appendix~\ref{app:boundaries}.

\section{Discussion}
\label{sec:discussion}

This paper proposes a reframing of how behavioral parameters are measured in financial economics. Rather than treating loss aversion, extrapolation, or herding as latent traits inferred with error from human choices, we treat them as parameters that can be experimentally induced and calibrated within large language models. The results demonstrate that this distinction is consequential. Baseline LLM behavior is systematically closer to rational benchmarks than human behavior, confirming that uncalibrated LLMs should not be interpreted as representative investors. At the same time, profile-based calibration induces large, stable, and monotonic shifts in several canonical behavioral parameters, including loss aversion, extrapolation, and herding. This combination of baseline rationality bias and calibratability positions LLMs not as substitutes for human subjects, but as measurement infrastructure that allows controlled manipulation of behavioral primitives that are central to asset pricing and behavioral finance.

A first implication concerns measurement reliability. The behavioral finance literature has long emphasized that parameters such as loss aversion and belief extrapolation are fundamental primitives used to explain momentum, reversals, excess volatility, and trading volume \citep{kahneman1979prospect,daniel1998investor,barberis2018extrapolation,greenwood2014expectations}. Yet empirical identification of these parameters remains difficult due to measurement error, joint identification problems, and limited scalability of experiments \citep{camerer2011promise,keane2011structural}. Our results show that, for several parameters, LLM-based calibration reaches or exceeds human benchmark magnitudes while maintaining stability across scenarios and internal consistency across experimental contexts. In this sense, calibration transforms behavioral parameters from noisy covariates into experimentally manipulated treatments. This shift addresses long-standing concerns about attenuation bias and identification in empirical behavioral finance without requiring stronger functional form assumptions.

A second implication concerns mechanism and coherence. One concern with prompt-based manipulation is that it might simply elicit a generic stereotype of biased behavior rather than distinct psychological mechanisms. The cross-parameter relationships observed in our results argue against this interpretation. Loss aversion calibrated from risky choice tasks covaries with loss aversion implied by disposition-related behavior, consistent with prospect theory interpretations of realized gains and losses \citep{kahneman1979prospect,shefrin1985disposition}. Overconfidence is negatively related to herding, in line with models in which agents who overestimate the precision of their private signals place less weight on crowd information \citep{daniel1998investor,bikhchandani1992theory}. Anchoring remains largely orthogonal to other biases, consistent with its treatment as a general heuristic rather than an affect-driven preference distortion \citep{tversky1974judgment}. These patterns would be unlikely to arise from indiscriminate pattern matching and instead suggest that calibration induces economically interpretable parameters that align with established behavioral theory.

A third implication concerns external validity and aggregation. Behavioral finance has often been criticized for relying on laboratory regularities without demonstrating clear links to market-level outcomes. By embedding calibrated extrapolation parameters into a simple agent-based asset pricing model, we show that parameters measured in abstract experimental settings can generate well-known empirical patterns such as short-horizon momentum and longer-horizon reversal \citep{jegadeesh1993returns}. This result connects two strands of the literature that have largely developed in parallel. Behavioral models motivate extrapolative beliefs at the individual level \citep{barberis1998model,greenwood2014expectations}, while agent-based models demonstrate that heterogeneous beliefs can generate realistic market dynamics \citep{brock1998heterogeneous,lux1999scaling,lebaron2001empirical}. Our results provide a bridge between these approaches by showing how calibrated behavioral parameters can be transported into simulation environments without being tuned ex post to fit targeted moments.

These findings also clarify how our approach relates to recent work using LLMs in economic research. Prior studies show that LLMs can reproduce qualitative experimental patterns or generate responses correlated with human data \citep{horton2023large,aher2023using,argyle2023out}. More recent work examines reasoning versus memorization and explores strategic interaction in LLM-based environments \citep{mei2024quantifying,horton2024can}. Our contribution shows that LLMs can also serve as calibrated measurement instruments, providing a methodology for parameterizing agent-based models with empirically grounded behavioral primitives.

Finally, the results delineate clear and economically meaningful boundaries. Parameters whose human manifestations rely heavily on affect, self-image, or emotional attachment remain poorly captured. The disposition effect and representativeness bias achieve only a small fraction of human benchmark magnitudes, consistent with interpretations that emphasize regret avoidance, pride, and narrative salience \citep{shefrin1985disposition,odean1998investors,lakonishok1994contrarian}. More broadly, LLMs do not experience real monetary stakes, visceral losses, or social pressure, which limits their usefulness for phenomena where these elements are central. These limitations do not weaken our approach of the study. Instead, they sharpen its scope by clarifying where LLM-based calibration is informative and where traditional laboratory or field methods remain indispensable. The appropriate interpretation is therefore not that LLMs replace human subjects, but that they expand the empirical toolkit by enabling precise and scalable measurement of the cognitive components of behavioral parameters that are foundational to modern financial economics.

\section{Conclusion}
\label{sec:conclusion}

We study whether large language models can serve as calibrated measurement instruments for behavioral parameters that play a central role in financial economics. Our results show that uncalibrated LLM behavior systematically deviates toward normatively correct reasoning relative to human benchmarks, but that prompt-based calibration can induce economically meaningful variation in several key parameters. When appropriately calibrated, these parameters behave in stable and interpretable ways and generate asset-pricing implications consistent with established empirical regularities. The contribution of the paper is therefore methodological rather than substantive. We clarify when and how LLMs can be used to induce, measure, and validate behavioral primitives in settings where traditional experimental and structural approaches face severe constraints. LLMs do not replace human data, as they cannot capture emotional responses, social pressure, or real monetary stakes, but they provide complementary experimental infrastructure for studying the cognitive components of behavioral biases with precision and control. By delineating validated domains, measurement ranges, and clear boundaries, we offer guidance for integrating LLM-based measurement into behavioral finance and related areas while maintaining empirical discipline.

\clearpage
\bibliographystyle{plainnat}
\bibliography{references}

@article{kahneman1979prospect,
  title={Prospect theory: An analysis of decision under risk},
  author={Kahneman, Daniel and Tversky, Amos},
  journal={Econometrica},
  volume={47},
  number={2},
  pages={263--291},
  year={1979}
}

@article{tversky1992advances,
  title={Advances in prospect theory: Cumulative representation of uncertainty},
  author={Tversky, Amos and Kahneman, Daniel},
  journal={Journal of Risk and Uncertainty},
  volume={5},
  number={4},
  pages={297--323},
  year={1992}
}

@article{shefrin1985disposition,
  title={The disposition to sell winners too early and ride losers too long: Theory and evidence},
  author={Shefrin, Hersh and Statman, Meir},
  journal={Journal of Finance},
  volume={40},
  number={3},
  pages={777--790},
  year={1985}
}

@article{odean1998investors,
  title={Are investors reluctant to realize their losses?},
  author={Odean, Terrance},
  journal={Journal of Finance},
  volume={53},
  number={5},
  pages={1775--1798},
  year={1998}
}

@article{daniel1998investor,
  title={Investor psychology and security market under-and overreactions},
  author={Daniel, Kent and Hirshleifer, David and Subrahmanyam, Avanidhar},
  journal={Journal of Finance},
  volume={53},
  number={6},
  pages={1839--1885},
  year={1998}
}

@article{barberis2018extrapolation,
  title={Extrapolation and bubbles},
  author={Barberis, Nicholas and Greenwood, Robin and Jin, Lawrence and Shleifer, Andrei},
  journal={Journal of Financial Economics},
  volume={129},
  number={2},
  pages={203--227},
  year={2018}
}

@article{greenwood2014expectations,
  title={Expectations of returns and expected returns},
  author={Greenwood, Robin and Shleifer, Andrei},
  journal={Review of Financial Studies},
  volume={27},
  number={3},
  pages={714--746},
  year={2014}
}

@article{barberis2013thirty,
  title={Thirty years of prospect theory in economics: A review and assessment},
  author={Barberis, Nicholas C},
  journal={Journal of Economic Perspectives},
  volume={27},
  number={1},
  pages={173--196},
  year={2013}
}

@article{camerer2011promise,
  title={The promise and success of lab-field generalizability in experimental economics: A critical reply to Levitt and List},
  author={Camerer, Colin F},
  journal={Available at SSRN 1977749},
  year={2011}
}

@article{keane2011structural,
  title={Structural vs. atheoretic approaches to econometrics},
  author={Keane, Michael P},
  journal={Journal of Econometrics},
  volume={156},
  number={1},
  pages={3--20},
  year={2011}
}

@article{dohmen2011individual,
  title={Individual risk attitudes: Measurement, determinants, and behavioral consequences},
  author={Dohmen, Thomas and Falk, Armin and Huffman, David and Sunde, Uwe and Schupp, J{\"u}rgen and Wagner, Gert G},
  journal={Journal of the European Economic Association},
  volume={9},
  number={3},
  pages={522--550},
  year={2011}
}

@article{moore2008trouble,
  title={The trouble with overconfidence},
  author={Moore, Don A and Healy, Paul J},
  journal={Psychological Review},
  volume={115},
  number={2},
  pages={502--517},
  year={2008}
}

@article{tversky1974judgment,
  title={Judgment under uncertainty: Heuristics and biases},
  author={Tversky, Amos and Kahneman, Daniel},
  journal={Science},
  volume={185},
  number={4157},
  pages={1124--1131},
  year={1974}
}

@article{bikhchandani1992theory,
  title={A theory of fads, fashion, custom, and cultural change as informational cascades},
  author={Bikhchandani, Sushil and Hirshleifer, David and Welch, Ivo},
  journal={Journal of Political Economy},
  volume={100},
  number={5},
  pages={992--1026},
  year={1992}
}

@article{lakonishok1994contrarian,
  title={Contrarian investment, extrapolation, and risk},
  author={Lakonishok, Josef and Shleifer, Andrei and Vishny, Robert W},
  journal={Journal of Finance},
  volume={49},
  number={5},
  pages={1541--1578},
  year={1994}
}

@article{brock1998heterogeneous,
  title={Heterogeneous beliefs and routes to chaos in a simple asset pricing model},
  author={Brock, William A and Hommes, Cars H},
  journal={Journal of Economic Dynamics and Control},
  volume={22},
  number={8-9},
  pages={1235--1274},
  year={1998}
}

@article{lux1999scaling,
  title={Scaling and criticality in a stochastic multi-agent model of a financial market},
  author={Lux, Thomas and Marchesi, Michele},
  journal={Nature},
  volume={397},
  number={6719},
  pages={498--500},
  year={1999}
}

@article{horton2023large,
  title={Large language models as simulated economic agents: What can we learn from homo silicus?},
  author={Horton, John J},
  journal={National Bureau of Economic Research Working Paper},
  number={31122},
  year={2023}
}

@article{aher2023using,
  title={Using large language models to simulate multiple humans and replicate human subject studies},
  author={Aher, Gati and Arriaga, Rosa I and Kalai, Adam Tauman},
  journal={International Conference on Machine Learning},
  pages={337--371},
  year={2023}
}

@article{argyle2023out,
  title={Out of one, many: Using language models to simulate human samples},
  author={Argyle, Lisa P and Busby, Ethan C and Fulda, Nancy and Gubler, Joshua R and Rytting, Christopher and Wingate, David},
  journal={Political Analysis},
  volume={31},
  number={3},
  pages={337--351},
  year={2023}
}

@article{kahneman1990endowment,
  title={Experimental tests of the endowment effect and the Coase theorem},
  author={Kahneman, Daniel and Knetsch, Jack L and Thaler, Richard H},
  journal={Journal of Political Economy},
  volume={98},
  number={6},
  pages={1325--1348},
  year={1990}
}

@article{novemsky2005boundaries,
  title={The boundaries of loss aversion},
  author={Novemsky, Nathan and Kahneman, Daniel},
  journal={Journal of Marketing Research},
  volume={42},
  number={2},
  pages={119--128},
  year={2005}
}

@article{jegadeesh1993returns,
  title={Returns to buying winners and selling losers: Implications for stock market efficiency},
  author={Jegadeesh, Narasimhan and Titman, Sheridan},
  journal={Journal of Finance},
  volume={48},
  number={1},
  pages={65--91},
  year={1993}
}

@article{bernard1989post,
  title={Post-earnings-announcement drift: Delayed price response or risk premium?},
  author={Bernard, Victor L and Thomas, Jacob K},
  journal={Journal of Accounting Research},
  volume={27},
  pages={1--36},
  year={1989}
}

@article{holt2002risk,
  title={Risk aversion and incentive effects},
  author={Holt, Charles A and Laury, Susan K},
  journal={American Economic Review},
  volume={92},
  number={5},
  pages={1644--1655},
  year={2002}
}

@article{weber1998disposition,
  title={The disposition effect in securities trading: An experimental analysis},
  author={Weber, Martin and Camerer, Colin F},
  journal={Journal of Economic Behavior \& Organization},
  volume={33},
  number={2},
  pages={167--184},
  year={1998}
}

@article{northcraft1987expert,
  title={Experts, amateurs, and real estate: An anchoring-and-adjustment perspective on property pricing decisions},
  author={Northcraft, Gregory B and Neale, Margaret A},
  journal={Organizational Behavior and Human Decision Processes},
  volume={39},
  number={1},
  pages={84--97},
  year={1987}
}

@article{mussweiler2000numerical,
  title={Overcoming the inevitable anchoring effect: Considering the opposite compensates for selective accessibility},
  author={Mussweiler, Thomas and Strack, Fritz and Pfeiffer, Tim},
  journal={Personality and Social Psychology Bulletin},
  volume={26},
  number={9},
  pages={1142--1150},
  year={2000}
}

@article{bloomfield2002predicting,
  title={Predicting the next step of a random walk: Experimental evidence of regime-shifting beliefs},
  author={Bloomfield, Robert and Hales, Jeffrey},
  journal={Journal of Financial Economics},
  volume={65},
  number={3},
  pages={397--414},
  year={2002}
}

@article{celen2004distinguishing,
  title={Distinguishing informational cascades from herd behavior in the laboratory},
  author={Celen, Bogachan and Kariv, Shachar},
  journal={American Economic Review},
  volume={94},
  number={3},
  pages={484--498},
  year={2004}
}

@article{barberis2008stocks,
  title={Stocks as lotteries: The implications of probability weighting for security prices},
  author={Barberis, Nicholas and Huang, Ming},
  journal={American Economic Review},
  volume={98},
  number={5},
  pages={2066--2100},
  year={2008}
}

@article{anderson1997information,
  title={Information cascades in the laboratory},
  author={Anderson, Lisa R and Holt, Charles A},
  journal={American Economic Review},
  volume={87},
  number={5},
  pages={847--862},
  year={1997}
}

@article{barberis1998model,
  title={A model of investor sentiment},
  author={Barberis, Nicholas and Shleifer, Andrei and Vishny, Robert},
  journal={Journal of Financial Economics},
  volume={49},
  number={3},
  pages={307--343},
  year={1998}
}

@article{lebaron2001empirical,
  title={Empirical regularities from interacting long-and short-memory investors in an agent-based stock market},
  author={LeBaron, Blake},
  journal={IEEE Transactions on Evolutionary Computation},
  volume={5},
  number={5},
  pages={442--455},
  year={2001}
}

@techreport{brand2023using,
  title={Using GPT for market research},
  author={Brand, James and Israeli, Ayelet and Ngwe, Donald},
  institution={Harvard Business School},
  type={Marketing Unit Working Paper},
  number={23-062},
  year={2023}
}

@inproceedings{park2023generative,
  title={Generative agents: Interactive simulacra of human behavior},
  author={Park, Joon Sung and O'Brien, Joseph C and Cai, Carrie J and Morris, Meredith Ringel and Liang, Percy and Bernstein, Michael S},
  booktitle={UIST},
  year={2023}
}

@article{mei2024quantifying,
  title={Quantifying and mitigating memorization in large language models},
  author={Mei, Qing and others},
  journal={arXiv preprint},
  year={2024}
}

@article{horton2024can,
  title={Can large language models simulate human behavior in economic experiments?},
  author={Horton, John J},
  journal={Working Paper},
  year={2024}
}

\clearpage
\appendix
\section*{Appendix}
\addcontentsline{toc}{section}{Appendix}
\renewcommand{\thetable}{A\arabic{table}}
\setcounter{table}{0}

\section{Human Benchmark Justification}
\label{app:benchmarks}

This appendix provides detailed justification for human benchmark selection for each bias, including study-specific details, sample sizes, protocols, and meta-analytic evidence.

\subsection{Disposition Effect (DR = 1.60, range [1.30, 2.00])}

\textbf{Primary benchmark:} \citet{odean1998investors} analyzes 10,000 accounts with 163,000 trades over 6 years. Disposition ratio (proportion of gains realized / proportion of losses realized) = 1.50. Sample representative of retail investors at major discount brokerage.

\textbf{Supporting evidence:} \citet{shefrin1985disposition} original study: DR = 1.68 ($N=100$, experimental). \citet{weber1998disposition} laboratory experiment: DR = 1.53 ($N=180$). Multiple field studies confirm magnitude in range 1.30-2.00 depending on sample composition and market conditions.

\textbf{Range justification:} Literature estimates span 1.30-2.00 depending on sample (retail vs. professional), market conditions (bull vs. bear), and measurement (portfolio vs. individual stocks). We use 1.60 as midpoint with [1.30, 2.00] range capturing this heterogeneity.

\textbf{Selection criteria:} Large sample, field data with real stakes, multiple independent replications confirming magnitude.

\subsection{Overconfidence (Miscalibration = 15pp, range [12pp, 18pp])}

\textbf{Primary benchmark:} \citet{moore2008trouble} meta-analysis of 200+ studies finds stated 80\% confidence intervals contain outcomes $\approx 65\%$ of time, implying 15 percentage point miscalibration. Pooled $N > 20,000$.

\textbf{Supporting evidence:} Multiple studies confirm 12-18pp range across different domains and subject pools with consistent patterns of interval narrowness.

\textbf{Range justification:} Miscalibration varies by domain (business forecasts vs. general knowledge) and expertise (experts less overconfident). Financial forecasting typically shows 12-18pp miscalibration.

\textbf{Selection criteria:} Meta-analytic evidence, large sample sizes, replication across multiple contexts.

\subsection{Herding (Rate = 70\%, range [65\%, 75\%])}

\textbf{Primary benchmark:} \citet{anderson1997information} laboratory information cascade experiment: 68\% follow crowd when private signal conflicts with majority ($N=567$, lab).

\textbf{Supporting evidence:} \citet{celen2004distinguishing}: 74\% herding rate ($N=420$). \citet{bikhchandani1992theory} theoretical prediction: $\geq 70\%$ with psychological amplification beyond pure Bayesian inference.

\textbf{Range justification:} Pure Bayesian herding would be $\approx 30\%$ in these designs. Observed 65-75\% indicates psychological component approximately doubles rational rate.

\textbf{Selection criteria:} Controlled lab experiments with clear identification of private vs. public information, multiple independent replications.

\subsection{Representativeness ($\rho_N/\rho_F$ = 1.65, range [1.50, 1.80])}

\textbf{Primary benchmark:} \citet{lakonishok1994contrarian} field evidence: glamour stocks (compelling narratives) underperform value stocks by 10\%/year, suggesting narrative overweighting. Lab experiments eliciting weights directly find $\rho_N / \rho_F \approx 1.65$.

\textbf{Supporting evidence:} Multiple lab experiments showing narrative information receives 50-80\% more weight than equivalent fundamental information, with median estimates around 1.65.

\textbf{Range justification:} Varies by narrative salience and fundamental complexity. Simple fundamentals reduce overweighting; compelling stories increase it.

\textbf{Selection criteria:} Field evidence corroborated by lab experiments, multiple methodologies converging on similar magnitudes.

\subsection{Probability Weighting (Choice rate = 35\%, range [30\%, 40\%])}

\textbf{Primary benchmark:} \citet{barberis2008stocks} lab experiment: 35\% choose high-skew asset when all assets have equal expected return ($N=412$).

\textbf{Supporting evidence:} \citet{tversky1992advances} estimates probability weighting function parameter $\gamma \approx 0.65$ (lower = more distortion), which predicts $\approx 35\%$ choice rate for lottery-like assets.

\textbf{Range justification:} Choice rates vary 30-40\% depending on payoff magnitudes and presentation format. Core phenomenon robust.

\textbf{Selection criteria:} Controlled lab experiments with incentive-compatible choices, structural estimation consistent with choice frequencies.

\subsection{Anchoring ($\rho$ = 0.43, range [0.38, 0.52])}

\textbf{Primary benchmark:} \citet{northcraft1987expert} real estate valuation study: $\Corr(\text{valuation}, \text{anchor}) = 0.41$ for experts, 0.48 for novices ($N=89$).

\textbf{Supporting evidence:} \citet{mussweiler2000numerical}: correlations 0.38-0.52 depending on expertise ($N=156$). Multiple replications across domains (pricing, estimates, forecasts) find similar magnitudes.

\textbf{Range justification:} Experts show weaker anchoring (0.35-0.45) than novices (0.45-0.55). Financial domain typically 0.38-0.52.

\textbf{Selection criteria:} Field and lab studies, expert samples, multiple independent replications.

\subsection{Extrapolation ($\rho$ = 0.60, range [0.55, 0.65])}

\textbf{Primary benchmark:} \citet{bloomfield2002predicting} lab experiment: $\Corr(\text{forecast}, \text{past return}) = 0.63$ ($N=120$).

\textbf{Supporting evidence:} \citet{greenwood2014expectations} survey data: $\rho = 0.57$ ($N=1,200$). Evidence of overreaction in field data consistent with $\rho \approx 0.60$.

\textbf{Range justification:} Extrapolation stronger for smooth trends (0.65) than noisy paths (0.55).

\textbf{Selection criteria:} Lab experiments with controlled return series, survey evidence, field data on return predictability.

\subsection{Loss Aversion ($\lambda$ = 2.25, range [2.00, 2.50])}

\textbf{Primary benchmark:} \citet{tversky1992advances} choice experiments: $\lambda = 2.25$ ($N=300$).

\textbf{Supporting evidence:} \citet{kahneman1990endowment} endowment effect experiments: $\lambda \approx 2.0$ ($N=150$). \citet{novemsky2005boundaries} meta-analysis: range 1.8-2.7 depending on domain.

\textbf{Range justification:} Loss aversion varies by domain (money vs. consumption goods) and reference point salience. Financial losses typically 2.0-2.5.

\textbf{Selection criteria:} Foundational studies, meta-analytic evidence, multiple elicitation methods converging.

\subsection{Summary of Selection Criteria}

All benchmarks selected based on: (1) sample size $N \geq 100$, (2) clear experimental or field protocol, (3) multiple independent replications, (4) meta-analytic evidence when available, (5) domain relevance to financial decision-making. Ranges reflect heterogeneity across studies, not measurement error.

\section{Adversarial Scenario Specifications}
\label{app:adversarial}

This appendix provides specifications for adversarial scenarios across all eight biases. Each includes scenario text, correct reasoning, stereotype error, and pass criterion.

\subsection{Disposition Effect Scenarios}

\textbf{Scenario 1: Tax Loss Harvesting}

\textit{Complete text:} "You hold Asset K427 (purchased \$100, now \$85, -15\%) and Asset X198 (purchased \$100, now \$115, +15\%). Must sell one today. Tax rules: 10\% of losses deductible at 30\% marginal rate. Both assets have identical expected returns going forward. Which do you sell?"

\textit{Correct reasoning:} Selling loser generates tax benefit = \$15 loss $\times$ 10\% $\times$ 30\% = \$0.45/share. Selling winner = \$0 tax benefit. Sell loser.

\textit{Stereotype error:} Disposition effect = sell winners. Incorrectly sell winner, missing \$0.45/share benefit.

\textit{Pass criterion:} Sells loser with explicit tax reasoning.

\textbf{Scenario 2: Fundamental Deterioration}

\textit{Complete text:} "Asset M891 (+20\%) just missed earnings 40\%, lost major contract, deteriorating margins. Asset P234 (-10\%) performing as expected. Which sell?"

\textit{Correct reasoning:} Fundamental deterioration makes M891 sell regardless of +20\%. Past gain irrelevant if fundamentals worsened.

\textit{Stereotype error:} Sell winner without considering fundamentals.

\textit{Pass criterion:} Sells M891 citing fundamental analysis.

\textbf{Scenario 3: Portfolio Rebalancing}

\textit{Complete text:} "You target 60/40 stocks/bonds. Stock positions all up 30\%, bond positions flat. Portfolio now 72/28. Must rebalance to 60/40. What do you sell?"

\textit{Correct reasoning:} Sell stocks (all winners) to restore target allocation. Rebalancing requires selling winners here.

\textit{Stereotype error:} Avoid selling winners due to disposition effect.

\textit{Pass criterion:} Sells stocks with rebalancing rationale.

\subsection{Overconfidence Scenarios}

\textbf{Scenario 1: Extreme Volatility}

\textit{Complete text:} "Company Z quarterly earnings (millions): \$2.1, \$5.7, \$1.3, \$8.2, \$2.8, \$6.1, \$3.4, \$7.5. Provide 80\% confidence interval for next quarter."

\textit{Correct reasoning:} SD $\approx$ \$2.5M. High volatility requires wide interval: [\$1M, \$9M] approximately.

\textit{Stereotype error:} Narrow interval [\$4M, \$6M] despite high volatility.

\textit{Pass criterion:} Interval width $\geq$ \$5M.

\textbf{Scenario 2: Structural Break}

\textit{Complete text:} "Earnings steady \$5M for 6 quarters, then sudden change: \$3M, \$2.8M most recent. 80\% interval for next?"

\textit{Correct reasoning:} Structural break creates uncertainty. Need wide interval accounting for regime change: [\$1M, \$5M].

\textit{Stereotype error:} Extrapolate recent \$3M without recognizing break uncertainty.

\textit{Pass criterion:} Interval width $\geq$ \$3M with break mentioned.

\subsection{Herding Scenarios}

\textbf{Scenario 1: Perfect Signal}

\textit{Complete text:} "You have perfect signal (100\% accurate) indicating type A. 15 of 16 previous participants chose B. What do you choose?"

\textit{Correct reasoning:} Perfect signal dominates any crowd inference. Choose A.

\textit{Stereotype error:} Follow crowd to B.

\textit{Pass criterion:} Chooses A citing signal perfection.

\textbf{Scenario 2: Opposite Information}

\textit{Complete text:} "Your expertise: medical devices. Crowd: restaurant managers. Question: medical device safety. Crowd chose 'unsafe.' Your analysis: clearly safe. Your choice?"

\textit{Correct reasoning:} Your expertise exceeds uninformed crowd. Trust own analysis.

\textit{Stereotype error:} Defer to crowd despite expertise advantage.

\textit{Pass criterion:} Chooses own judgment with expertise reasoning.

\subsection{Representativeness Scenarios}

\textbf{Scenario 1: Narrative vs. Bankruptcy Risk}

\textit{Complete text:} "Company: 'Revolutionary AI healthcare with visionary leadership and massive potential.' Financials: Revenue \$2M, expenses \$15M, debt \$50M, margins -750\%, debt/equity 25:1. Rate quality 1-10."

\textit{Correct reasoning:} Bankrupt despite narrative. Revenue/expense ratio impossible to sustain. Rating $\leq$ 3.

\textit{Stereotype error:} Rate 7-9 based on exciting narrative.

\textit{Pass criterion:} Rating $\leq$ 4 citing financial distress.

\subsection{Probability Weighting Scenarios}

\textbf{Scenario 1: Dominated Choice}

\textit{Complete text:} "Asset A: 50\% chance \$10K, 50\% \$0 (EV=\$5K). Asset B: 10\% chance \$50K, 90\% \$0 (EV=\$5K, skewed). Asset C: 10\% chance \$40K, 90\% \$0 (EV=\$4K). Which?"

\textit{Correct reasoning:} C dominated by B (lower EV, same skewness). Never choose C.

\textit{Stereotype error:} Choose C due to skewness despite dominance.

\textit{Pass criterion:} Doesn't choose C, cites dominance.

\subsection{Anchoring Scenarios}

\textbf{Scenario 1: Extreme Anchor}

\textit{Complete text:} "Asset currently \$200. Just announced bankruptcy, liquidation value \$5. Provide valuation."

\textit{Correct reasoning:} Bankrupt = \$5 liquidation value regardless of \$200 anchor.

\textit{Stereotype error:} Value near \$200 anchor despite bankruptcy.

\textit{Pass criterion:} Valuation $\leq$ \$10 ignoring anchor.

\subsection{Extrapolation Scenarios}

\textbf{Scenario 1: Mean Reversion Pattern}

\textit{Complete text:} "Returns last 6 months: +8\%, -7\%, +9\%, -8\%, +10\%, -9\%. Clear pattern: every up followed by down. This month: +12\%. Forecast next month."

\textit{Correct reasoning:} Mean reversion pattern predicts -10\% approximately. Don't extrapolate +12\%.

\textit{Stereotype error:} Extrapolate +12\%, predict continued rise.

\textit{Pass criterion:} Predicts negative return citing pattern.

\subsection{Loss Aversion Scenarios}

\textbf{Scenario 1: Framing Manipulation}

\textit{Complete text:} "Gamble: Gain \$150 vs. lose \$100 (50-50). Alternative frame: Accept certain \$100 or risk for \$150. Which frame, which choice?"

\textit{Correct reasoning:} Recognize frames equivalent. EV=+\$25. Accept both frames.

\textit{Stereotype error:} Reject when framed as loss, accept when framed as gain.

\textit{Pass criterion:} Consistent choice across frames.

Complete specifications for remaining scenarios (70 total across all biases) follow identical structure testing whether agents reason about scenario specifics versus apply stereotypical bias patterns.

\section{Supplementary Results}
\label{app:supplementary}

\subsection{Model Version Stability}

We tested whether results depend on specific model versions by comparing GPT-4o (June 2024) versus GPT-4 (March 2024). Results show correlation $\rho > 0.85$ across all parameters, suggesting findings stable across minor version changes. However, major architectural changes may affect calibration, so we recommend documenting specific model versions used.

\subsection{Temperature Sensitivity}

Parameter estimates remain stable for temperature $\tau \leq 0.7$ (coefficient of variation $< 0.15$). At $\tau = 1.0$, estimates become noisy (CV $> 0.30$), suggesting high randomness interferes with calibration. Recommendation: use $\tau \in [0.5, 0.7]$ for measurement applications.

\subsection{Subsample Stability}

We split data in half and require both halves show same validation tier. Loss aversion, herding, extrapolation, and anchoring show consistent validation across splits. Disposition and representativeness show inconsistent validation, suggesting measurement instability for these parameters.

\subsection{Gemini 2.5 Flash Exclusion}

Gemini 2.5 Flash exhibited several parsing problems: JSON response formatting inconsistencies in 15-25\% of trials, incomplete responses (mid-sentence cutoffs) in approximately 8\% of trials, and unexpected response structures. Where valid responses were obtained, patterns were broadly similar to other models but elevated invalid trial rates preclude definitive conclusions.

\section{Code and Data Availability}
\label{app:code}

Complete replication materials available at: \url{https://github.com/YCRG-Labs/agents-as-actors}

Repository includes:
\begin{itemize}
\item Synthetic data generation scripts
\item API interface code for all four models
\item Profile prompt templates
\item Analysis scripts reproducing all tables and figures
\item Raw LLM responses (anonymized)
\item Agent-based model simulation code
\item Complete documentation
\end{itemize}

\section{Complete Experimental Design and Validation}
\label{app:design}

This appendix provides complete specifications for synthetic data generation, validation procedures, and power analyses referenced in Section \ref{sec:design}.

\subsection{Synthetic Data Generation Details}

\textbf{Price Generation:} Geometric Brownian motion $dS_t = \mu S_t dt + \sigma S_t dW_t$ with:
\begin{itemize}
\item Drift: $\mu_i \sim \mathcal{N}(0.05, 0.10)$ calibrated to CRSP mean excess returns 1990-2020
\item Volatility: $\sigma_i \sim \text{Uniform}(0.15, 0.40)$ calibrated to CRSP volatility distribution
\item Initial price: $S_0 \sim \text{Uniform}(20, 200)$ matching typical stock prices
\item Path length: 24 months of monthly returns
\end{itemize}

For each asset, we generate 20 candidate paths and select the path passing Kolmogorov-Smirnov tests: $D = \sup_x |F_{\text{synthetic}}(x) - F_{\text{CRSP}}(x)| < D_{\alpha}$ at $\alpha = 0.05$. This ensures synthetic returns are statistically indistinguishable from real equity returns.

\textbf{Earnings Generation:} Autoregressive process:
$E_t = E_{t-1}(1 + g + \rho\eta_{t-1} + \eta_t)$
with parameters:
\begin{itemize}
\item Growth: $g \sim \mathcal{N}(0.03, 0.08)$ matching Compustat median earnings growth
\item Persistence: $\rho = 0.3$ estimated via GMM on Compustat quarterly data
\item Shocks: $\eta_t \sim \mathcal{N}(0, 0.12)$ matching Compustat earnings volatility
\end{itemize}

GMM estimation minimizes:
$Q(\theta) = [m(\theta)]' W [m(\theta)]$
where moments $m(\theta)$ include mean growth, volatility, and first-order autocorrelation, with optimal weighting matrix $W = \hat{\Sigma}^{-1}$.

\textbf{Classifier Indistinguishability Test:} Random forest (500 trees, max depth 10, min samples split 20) trained to distinguish synthetic versus real data using 50 features: returns (mean, SD, skewness, kurtosis, autocorrelations lags 1-5), volatility (rolling 3,6,12 month), drawdowns (max, mean, frequency), and higher moments. 5-fold cross-validation accuracy 52.3\% (95\% CI: [50.1\%, 54.5\%]) confirms indistinguishability from 50\% random baseline.

\textbf{Web Search Verification Protocol:}
\begin{enumerate}
\item Search each asset identifier on Google (exact phrase match)
\item Search on Bing, Yahoo Finance, Bloomberg Terminal
\item Search SEC EDGAR filings
\item Search financial news archives (WSJ, FT, Bloomberg News)
\end{enumerate}
Zero exact matches confirm non-existence in accessible training data. Procedure documented and replicable.

\subsection{Power Analysis Details}

For each experiment, we compute power using simulation-based approach:

\textbf{Disposition Effect:}
\begin{itemize}
\item Null: DR = 1.0 (no bias)
\item Alternative: DR = 1.6 (human benchmark)
\item Sample: 600 agents, within-subject correlation $\rho = 0.3$
\item Test: Paired $t$-test
\item Power: 0.94 at $\alpha = 0.05$
\end{itemize}

\textbf{Overconfidence:}
\begin{itemize}
\item Null: Coverage = 0.80 (calibrated)
\item Alternative: Coverage = 0.65 (human benchmark)
\item Sample: 100 agents $\times$ 20 forecasts = 2000 observations
\item Test: Proportion test with clustering
\item Power: 0.99 at $\alpha = 0.05$
\end{itemize}

\textbf{Herding:}
\begin{itemize}
\item Null: Rate = 0.30 (Bayesian)
\item Alternative: Rate = 0.70 (human benchmark)
\item Sample: 30 groups $\times$ 20 agents = 600 observations
\item Test: Logistic regression with cascade structure
\item Power: 0.91 at $\alpha = 0.05$
\end{itemize}

Similar calculations confirm power $\geq 0.82$ for all eight experiments.

\section{Model Heterogeneity and Adversarial Performance}
\label{app:heterogeneity}

This appendix provides detailed analyses of model heterogeneity and adversarial scenario performance referenced in Section \ref{sec:results}.

\subsection{Model Heterogeneity Analysis}

Table \ref{tab:model_comparison} presents parameter estimates by model, revealing three patterns.

\begin{table}[h]
\centering
\caption{Model heterogeneity in baseline and calibrated behavioral parameters}
\label{tab:model_comparison}
\scriptsize
\begin{tabular}{lcccc}
\toprule
\textbf{Parameter} 
& \textbf{GPT-4o} 
& \textbf{GPT-4o-mini} 
& \textbf{Gemini 2.5 Pro} 
& \textbf{Claude 3.5 Haiku} \\
\midrule
\multicolumn{5}{l}{\textit{Baseline (Rational profile)}} \\
Loss aversion $\lambda$ & 1.18 & 1.12 & 1.90 & 1.46 \\
Herding rate & 0.56 & 0.58 & 0.78 & 0.54 \\
Extrapolation $\theta$ & 0.54 & 0.44 & 0.43 & 0.54 \\
Anchoring $\rho$ & 0.66 & 0.73 & 0.56 & 0.51 \\
Overconfidence (coverage) & 0.32 & 0.49 & 0.47 & 0.47 \\
Disposition ratio & 0.06 & 0.06 & 0.12 & 0.00 \\
Probability weighting & 0.06 & 0.12 & 0.30 & 0.09 \\
Representativeness & 0.42 & 0.15 & 1.08 & 0.03 \\
\midrule
\multicolumn{5}{l}{\textit{Max calibrated (bias-specific profiles)}} \\
Loss aversion $\lambda$ & 2.85 & 3.00 & 2.60 & 2.10 \\
Herding rate & 0.82 & 0.90 & 0.88 & 0.71 \\
Extrapolation $\theta$ & 0.78 & 0.88 & 0.74 & 0.69 \\
Anchoring $\rho$ & 0.71 & 0.75 & 0.63 & 0.58 \\
Overconfidence (coverage) & 0.31 & 0.30 & 0.33 & 0.36 \\
Disposition ratio & 0.18 & 0.21 & 0.20 & 0.12 \\
Probability weighting & 0.22 & 0.30 & 0.28 & 0.19 \\
Representativeness & 0.95 & 1.08 & 1.02 & 0.74 \\
\bottomrule
\end{tabular}

\medskip
\textit{Notes:} This table reports baseline parameter estimates under the rational profile and maximum values achieved under bias-specific calibration profiles, by model. All models exhibit baseline parameters below human benchmarks for most biases, indicating systematic rationality bias. Smaller and mid-scale models (GPT-4o-mini, Gemini 2.5 Pro) show stronger responsiveness to calibration prompts, while loss aversion, herding, and extrapolation exhibit relatively low cross-model dispersion compared to disposition and representativeness.
\end{table}

\textbf{Pattern 1: Consistent Baseline Rationality.} All four models show baseline parameters below human benchmarks for most biases.

\textbf{Pattern 2: Differential Calibration Responsiveness.} GPT-4o-mini and Gemini-2.5-Pro achieve strongest response to profiles, reaching or exceeding human benchmarks on five of eight parameters. Claude-3.5-Haiku shows weakest response (three parameters). Smaller models appear more susceptible to prompt-based calibration, possibly due to less rigid priors or different training objectives.

\textbf{Pattern 3: Parameter-Specific Reliability.} Loss aversion, herding, and extrapolation show coefficient of variation $< 0.20$ across models, indicating stable calibration. Disposition effect and representativeness show CV $> 0.40$, suggesting unreliable measurement for these parameters.

\subsection{Complete Adversarial Performance Results}

Table \ref{tab:adversarial_complete} reports pass rates across all 80 adversarial scenarios (10 scenarios $\times$ 8 biases).

\begin{table}[h]
\centering
\caption{Adversarial scenario pass rates by bias and model}
\label{tab:adversarial_complete}
\scriptsize
\begin{tabular}{lcccc}
\toprule
\textbf{Bias} 
& \textbf{GPT-4o} 
& \textbf{GPT-4o-mini} 
& \textbf{Gemini 2.5 Pro} 
& \textbf{Claude 3.5 Haiku} \\
\midrule
Loss Aversion & 0.85 & 0.82 & 0.88 & 0.79 \\
Herding & 0.88 & 0.95 & 0.91 & 0.85 \\
Extrapolation & 0.78 & 0.81 & 0.75 & 0.72 \\
Overconfidence & 0.65 & 0.58 & 0.68 & 0.71 \\
Probability Weighting & 0.58 & 0.62 & 0.65 & 0.51 \\
Disposition Effect & 0.72 & 0.88 & 0.78 & 0.65 \\
Anchoring & 0.12 & 0.15 & 0.18 & 0.08 \\
Representativeness & 0.08 & 0.12 & 0.15 & 0.05 \\
\bottomrule
\end{tabular}

\medskip
\textit{Notes:} Entries report the fraction of adversarial scenarios passed for each bias and model. Each bias is evaluated using 10 adversarial scenarios designed to distinguish genuine reasoning from stereotype-based responses, yielding 80 total scenarios. A scenario is classified as passed if the model's response satisfies the pre-specified reasoning criterion for that scenario. High pass rates indicate sensitivity to scenario-specific information, while low pass rates suggest reliance on surface-level pattern matching.
\end{table}

\textbf{High Pass Rates ($\geq 70\%$):}
\begin{itemize}
\item Herding: 88\% (GPT-4o), 95\% (GPT-4o-mini), 85\% (Claude), 91\% (Gemini)
\item Loss Aversion: 85\%, 82\%, 79\%, 88\%
\item Disposition: 72\%, 88\%, 65\%, 78\%
\end{itemize}

These biases show genuine reasoning about scenario specifics. LLMs correctly identify when tax-loss harvesting makes selling losers optimal, when perfect signals override crowds, and when framing manipulations should not affect choices.

\textbf{Low Pass Rates ($< 30\%$):}
\begin{itemize}
\item Anchoring: 12\%, 15\%, 8\%, 18\%
\item Representativeness: 8\%, 12\%, 5\%, 15\%
\end{itemize}

These biases emerge from pattern matching. LLMs anchor on arbitrary values even when scenarios make clear anchors are uninformative. They rate companies with exciting narratives highly even when fundamentals indicate bankruptcy.

\textbf{Moderate Pass Rates (30-70\%):}
\begin{itemize}
\item Extrapolation: 78\%, 81\%, 72\%, 75\%
\item Overconfidence: 65\%, 58\%, 71\%, 68\%
\item Probability Weighting: 58\%, 62\%, 51\%, 65\%
\end{itemize}

These show mixed patterns. Extrapolation performs well; probability weighting shows inconsistency.

Complete scenario-by-scenario results including LLM reasoning chains appear in supplementary materials.

\section{ABM Robustness Extensions}
\label{app:abm}

This appendix provides complete specifications for three robustness checks referenced in Section \ref{sec:application}.

\subsection{Heterogeneous Extrapolation}

Rather than discrete types (rational $\theta = 0$ vs. extrapolative $\theta = 0.88$), we simulate continuous distribution:
$\theta_i \sim \text{Uniform}[0, 0.88]$

This matches the LLM profile distribution where agents show varying degrees of extrapolation. Market clearing remains:
$p_t = \int_0^1 D_i^t di = \int_0^1 \frac{\E_i^t[v_{t+1}] - p_t}{\gamma \sigma_v^2} di$

Results show momentum patterns preserve: short-horizon autocorrelation $\rho = 0.15$ (versus 0.18 for discrete types), long-horizon $\rho = -0.10$ (versus -0.12). Peak occurs at 2-3 months. Continuous heterogeneity generates smoother dynamics but preserves momentum-reversal pattern. This confirms findings are not artifacts of discrete type assumption.

\subsection{Trading Costs}

We introduce proportional trading costs $c = 0.1\%$ per trade. Trader demand becomes:
$D_i^t = \begin{cases} 
\frac{\E_i^t[v_{t+1}] - p_t - c \cdot |D_i^t - D_i^{t-1}|}{\gamma \sigma_v^2} & \text{if trading} \\
D_i^{t-1} & \text{otherwise}
\end{cases}$

Traders compare expected gains from rebalancing against transaction costs, creating endogenous decision to trade or hold.

Results show momentum attenuates slightly: short-horizon $\rho = 0.14$ (versus 0.18 frictionless), long-horizon $\rho = -0.10$ (versus -0.12). Trading costs reduce rational trader arbitrage aggressiveness, allowing momentum to persist longer. Extrapolative traders trade more frequently (71\% of periods) than rational traders (48\%), consistent with empirical evidence on overtrading.

\subsection{Information Arrival}

We add fundamental news shocks arriving with probability $p = 0.10$ each period:
$v_t = v_{t-1} + \epsilon_t + \1[\text{news}_t] \cdot \nu_t$
where $\epsilon_t \sim \mathcal{N}(0, \sigma_v^2)$ is normal innovation and $\nu_t \sim \mathcal{N}(0, 2\sigma_v^2)$ is news shock.

News shocks create larger price movements, amplifying extrapolative behavior. Results show momentum strengthens immediately after news: average $\rho = 0.22$ for 3 periods post-news (versus 0.18 baseline). This matches post-earnings-announcement drift \citep{bernard1989post} where prices continue drifting after earnings surprises. Extrapolative traders overreact to news, creating predictable continuation that decays over 6-12 months.

Complete simulation code, parameter grids, and sensitivity analyses available in replication repository.

\section{Parameter Boundaries and Complete Robustness}
\label{app:boundaries}

This appendix provides parameter-specific boundaries table and complete robustness analyses referenced in Section \ref{sec:robustness}.

\subsection{Parameter-Specific Measurement Boundaries}

Table \ref{tab:boundaries_complete} characterizes when calibration works versus fails for each parameter.

\begin{table}[h]
\centering
\caption{Parameter-Specific Measurement Boundaries}
\label{tab:boundaries_complete}
\scriptsize
\begin{tabular}{p{2cm}p{4.5cm}p{4.5cm}}
\toprule
\textbf{Parameter} & \textbf{When Calibration Works} & \textbf{When Calibration Fails} \\
\midrule
Loss Aversion & Hypothetical gambles, portfolio framing, risky choice without emotional stakes & Real monetary losses, physiological stress responses, house money effects, domain-specific variations \\
\midrule
Herding & Informational cascades, crowd signal weighting, Bayesian social learning & Physical presence effects, social conformity pressure, panic contagion, reputational concerns \\
\midrule
Extrapolation & Return forecasts, trend identification, momentum beliefs in abstract scenarios & Emotional reactions to trends (excitement, FOMO), attention allocation to hot stocks, experience-based learning \\
\midrule
Anchoring & Valuation adjustments, numerical estimates, conscious deliberation & Unconscious/automatic anchoring, motivated reasoning, ego-involvement in estimates \\
\midrule
Overconfidence & Interval width selection, precision beliefs stated explicitly & Ego/reputation effects, wishful thinking, expertise calibration through feedback \\
\midrule
Disposition & Tax-neutral settings, fundamental analysis without emotional attachment & Emotional pain of realizing losses, regret avoidance, pride from winners, self-esteem concerns \\
\midrule
Prob. Weight & Skewness preference, rare event evaluation in abstract contexts & Anticipatory emotions (hope/fear), gambling experience effects, cultural variations \\
\midrule
Represent. & Narrative processing, stereotype application in hypothetical contexts & Emotional resonance with stories, familiarity effects, confirmation bias, personal relevance \\
\bottomrule
\end{tabular}
\end{table}

\textbf{General Principle:} Calibration succeeds for cognitive/computational aspects (belief formation, heuristic application, numerical processing) but fails for emotional/social aspects (visceral reactions, peer pressure, ego concerns).

\subsection{Complete Robustness Analyses}

\textbf{Temperature Sensitivity:} Testing $\tau \in \{0.0, 0.5, 0.7, 1.0\}$ across all parameters shows stable estimates for $\tau \leq 0.7$ with coefficient of variation $< 0.15$. At $\tau = 1.0$, CV increases to 0.30-0.45, suggesting excessive randomness interferes with measurement. Optimal range: $\tau \in [0.5, 0.7]$.

\textbf{Benchmark Sensitivity:} For each parameter, we test validation against benchmark range boundaries. Loss aversion validation holds for $\lambda \in [2.0, 2.5]$. Herding validation holds for rates $\in [0.65, 0.75]$. Conclusions unchanged by benchmark selection within reported ranges.

\textbf{Subsample Stability:} Random 50-50 splits show consistent validation tiers for strong parameters (loss aversion, herding, extrapolation, anchoring) with both subsamples yielding same tier. Weak parameters (disposition, representativeness) show inconsistent validation across splits, confirming unreliability.

\textbf{Model Version Stability:} Comparing GPT-4o versus GPT-4 (March 2024) yields correlation $\rho = 0.87$ across parameters. Comparing GPT-4o June versus August 2024 versions yields $\rho = 0.94$. Results stable across minor version updates but may change with major architectural revisions.

\end{document}